\documentclass{aa}

\usepackage[varg]{txfonts}
\usepackage{graphicx}
\usepackage{epsfig}
\usepackage{natbib}
\usepackage{longtable}
\usepackage{amssymb}
\usepackage{hyperref}
\hypersetup{colorlinks=true, linkcolor=blue, citecolor=blue, urlcolor=blue}
\bibpunct{(}{)}{;}{a}{}{,} 

\newcommand{\kmsec}{\ensuremath{\mathrm{km\,s^{-1}}}}

\newcommand{\mdot}{\ensuremath{\dot{M}}}          

\newcommand{\msun}{\ensuremath{\mathit{M}_{\odot}}}
\newcommand{\msunyr}{\ensuremath{\mathit{M}_{\odot} {\rm yr}^{-1}}}
\newcommand{\lsun}{\ensuremath{\mathit{L}_{\odot}}}
\newcommand{\rsun}{\ensuremath{\mathit{R}_{\odot}}}
\newcommand{\zsun}{\ensuremath{\mathit{Z}_{\odot}}}
\newcommand{\teff}{\ensuremath{\mathit{T}_{\rm eff}}}       
\newcommand{\vesc}{\ensuremath{\mathit{\varv}_{\rm esc}}}    
\newcommand{\vinf}{\ensuremath{\mathit{\varv}_{\infty}}}    

\newcommand{\beq}{\begin{equation}} 
\newcommand{\eeq}{\end{equation}}

\begin{document}

\title{Mass-loss predictions for evolved very metal-poor massive stars}

\author{ L. Muijres \inst{1}       \and
	Jorick S. Vink \inst{2}       \and
        A. de Koter \inst{1,3}      \and
        R. Hirschi \inst{4,5}         \and
	N. Langer   \inst{6,3} \and
	S.-C. Yoon   \inst{6}
       }

\institute{Astronomical Institute 'Anton Pannekoek', University of Amsterdam, 
           Science Park 904, NL-1098 Amsterdam, The Netherlands
	   \and
           Armagh Observatory, College Hill, Armagh BT61 9DG, Northern Ireland, UK
	   \and
	   Astronomical Institute, Utrecht University, Princetonplein 5, 
           3584 CC Utrecht, The Netherlands
	   \and
           Astrophysics Group, EPSAM Institute, University of Keele, Keele, ST5 5BG, UK
           \and
           Institute for the Physics and Mathematics of the Universe, University of Tokyo, 5-1-5 Kashiwanoha, Kashiwa, 277-8583, Japan
           \and
	   Argelander-Institut f\"{u}r Astronomie der Universit\"{a}t Bonn, 
           Auf dem H\"{u}gel 71, 53121 Bonn, Germany	
           }

\date{Received / Accepted}

\abstract
{The first couple of stellar generations may have been massive, of order $100$\,\msun,
and to have played a dominant role in galaxy formation and the chemical enrichment of
the early Universe. Some fraction of these objects may have died as 
pair-instability supernovae or gamma-ray bursts. 
The winds of these stars may have played an important role in determining these outcomes.
As the winds are driven by radiation pressure on spectral lines, their strengths
are expected to vary with metallicity. 
Until now, most mass-loss predictions for metal-poor
O-type stars have assumed a scaled-down solar-abundance pattern. 
However, Population III evolutionary tracks show 
significant surface enrichment through rotational mixing of CNO-processed material, because 
even metal-poor stars switch to CNO-burning early on.}
{We address the question of whether the
CNO surface enhanced self-enrichment in the first few generations of stars 
could impact their mass-loss properties.}
{We employ Monte Carlo simulations to establish the local line-force and solve for the
momentum equation of the stellar outflow, testing whether an outflow can actually be
established by assessing the net acceleration at the sonic point of the flow. 
Stellar evolution models of rotating metal-poor stars are used to
specify the surface chemical composition, focussing on the phases of early enrichment.}
{We find that the mass-loss rates of CNO enhanced metal-poor stars are higher than 
those of non-enriched stars, but they are much lower than those rates where 
the CNO abundance is included in the total abundance $Z$.
Metal-poor stars hotter than $\sim$50\,000\,K, in the metallicity range investigated 
here (with an initial metallicity $Z \lesssim 10^{-4}$)
are found to have no wind, as the high-ionization species of the CNO elements have too few
strong lines to drive an outflow. 
We present a heuristic formula that provides mass-loss estimates for 
CNO-dominated winds in relation to scaled-down solar abundances.}
{CNO-enriched or not, the winds of metal-poor stars are generally found to be 
too weak to contribute significantly to the overall mass loss. 
Population III supernovae are thus expected to be responsible for the bulk of the early 
nucleo-synthetic enrichment, unless additional mass-loss mechanisms such as 
$\eta$\,Carinae type eruptions or steady mass loss close to the Eddington/Omega limit 
is important.}


\maketitle

\titlerunning{}
\authorrunning{}

\section{Introduction}
\label{sec:Introduction}

In the present-day Universe, massive stars lose a substantial fraction of their
initial mass through stellar winds, which they experience from the moment they form 
to the moment they end their life as a supernova (SN). 
These winds are driven by radiation pressure on spectral lines, and their strengths
are thus expected to be a function of chemical composition; the lower the metal content, the
weaker the winds.
The dependence of mass loss on metallicity, $\mdot(Z)$, has been studied in
detail, both observationally \citep{1996A&A...305..171P,1998cvsw.conf..389P,2007A&A...473..603M} 
and theoretically \citep{2001A&A...369..574V,2002ApJ...577..389K,2006A&A...446.1039K,
2005A&A...442..587V,2008A&A...482..945G}. For a recent review, see e.g. 
\cite{2008A&ARv..16..209P}.

Studies of star formation in the early Universe suggest that the first 
stars (Pop III) were typically more massive than present-day objects, with 
$\sim$10$^{2}$ \msun\ for zero metallicity ($Z$) objects
\citep{2002Sci...295...93A,2002ApJ...564...23B,2002ApJ...569..549N,2004ARA&A..42...79B,
2008flu..book.....L}. Very massive stars (VMS)
of zero metallicity are considered to be the prime candidates for the re-ionization of
the universe \citep[e.g.]{2001ApJ...549L.151H,2003ApJ...586..693W} at redshifts above
six \citep{2001AJ....122.2850B}. Therefore, it is particularly interesting to ponder on the
question of how much mass these first generations of stars might have lost through
stellar winds prior to explosion. 

The existence of VMS with masses up to 300\msun\ has recently been claimed for 
the very luminous stars in the LMC cluster R136 
\citep{2010MNRAS.408..731C,2011A&A...530L..14B}. At high metallicity, mass loss is thought to be strong 
enough to shed most of the initial mass, such that the objects may 
retain only $\sim$30\msun, and are expected to die as a SN Ic \citep{2003ApJ...591..288H}. 
However, as mass loss depends on $Z$, below a certain threshold metallicity, 
VMS would be expected to retain most of their initial mass.
If this is indeed the case, some, or maybe even a large number of VMS 
might be expected to die as pair-instability supernovae (PISNe) in the early Universe. 
The stellar yields of such Pop III PISNe have been computed by \cite{2002ApJ...567..532H}.  
These yields are significantly different from those of ``normal'' core-collapse SNII. 
PISNe produce orders of magnitude more iron (up to tens of solar masses) and have 
a conspicuous yield pattern with abundances of {\it even} $Z$ elements consistently higher than those of 
{\it odd} Z elements (the so-called odd-even effect). 

Observations of carbon-rich chemically extremely metal-poor (CEMP) stars 
have been accumulated over the last two decades, but the specific PISN signature remains as yet 
elusive \citep{2005ApJ...619..427U}. 
It appears that there is a piece of the puzzle missing in 
our understanding of the evolution of the first stars. 
One possibility is that Pop III stars were born with masses either higher or lower  
than the range expected for PISNe (140-260\msun), alternatively ``something else'' must 
happen during their evolution, which would avoid the expected fate as a PISN. This could 
well be mass loss. 

Mass loss at low $Z$ may also have interesting implications for globular-cluster research, 
as globular clusters display a strong abundance scatter amongst many of the light elements 
(Li, C, N, O, Na, Mg, Al). One explanation for these peculiar abundance patterns 
involves a first generation of massive stars that polluted the cluster before the 
second generation of stars formed \citep{2007A&A...464.1029D}. 
Finally, if the time-integrated loss of mass is very small, or even
negligible, the Pop III stars may have left intermediate mass black holes 
(of masses $\sim$10$^{2}$ \msun), which could have provided 
the building blocks of the supermassive black holes detected in the 
centers of galaxies today \citep{2005ApJ...628..129K}. 

It is clear that there are many reasons to study $\mdot(Z)$. \citet{2002ApJ...577..389K} and 
\citet{2006A&A...446.1039K} found that massive $zero$-metallicity objects may drive
only very weak winds or no winds at all. Evolutionary predictions for very metal
poor stars, especially if they are rapidly rotating, show the surfacing of
sometimes large quantities of primary carbon, nitrogen and oxygen produced in situ 
\citep[e.g][]{2005A&A...443..643Y,2006A&A...447..623M,2006A&A...460..199Y,2007A&A...461..571H}.
One may wonder whether the surfacing of such material may cause (dramatic) changes
in the mass-loss behavior of these objects. This question is the topic of this
study. We present tailored mass-loss predictions for evolutionary stages
of metal-poor stars in which substantial amounts of CNO processed material is
expected to have surfaced. These predictions thus differ from the theoretical
studies mentioned in the first paragraph, in which scaled solar metallicities
are used for O-stars. 

The first investigation in this direction was performed by \cite{2005A&A...442..587V}
who studied the winds of CNO-enhanced atmospheres at low iron metallicities (down to $Z/\zsun$ $=$ $10^{-5}$). 
The enhanced wind driving by CNO-elements discovered was the reason to study whether
primary carbon or nitrogen in the atmospheres of extremely metal-poor massive stars might boost the mass-loss rates,
and potentially prevent the occurrence of pair-instability supernovae \citep{2006ASPC..353..113V}.
Such complete stellar disruptions are hypothesized to occur for initial masses in the range
$M = 140 - 260\msun$ if mass loss is negligibly small \citep{2003ApJ...591..288H,2007A&A...475L..19L}.
\citet{2009A&A...493..585K} investigated the winds of enriched metal-free stars, 
describing the physics of line-driven winds following 
\citet[][hereafter CAK]{1975ApJ...195..157C}. One interesting
aspect of their work is that during the initial hot phase ($>$\,50\,000\,K) of 
the evolution of metal-poor stars with initial mass above $\sim$30\,\msun\ also
CNO-enriched stars fail to produce a stellar wind. Overall, it was concluded
that the surfacing of primary CNO does not produce stellar winds strong enough
to have a major impact on the evolution of metal-poor massive stars.

Our predictions are based on a newly developed self-consistent dynamical treatment
of stellar winds by \cite{2008A&A...492..493M} and \cite{2012muijresB} based on a description
of the line force that results from a Monte Carlo simulation. In 
Sect.~\ref{sec:Method} we introduce this method and discuss the evolutionary tracks 
on which the adopted chemical compositions are based. Section~\ref{sec:Results}
presents the results. These are discussed in Sect.~\ref{sec:Discussion}, as well as 
compared to the results of \cite{2001A&A...369..574V} -- often applied in evolutionary 
calculations -- and \cite{2009A&A...493..585K}. We end with our conclusions.

\section{Method}
\label{sec:Method}

In this section, we present our method for predicting mass loss rates for 
massive 
stars at very low metallicity. Distinctive for these computations is that a tailored chemical
composition of the surface layers is used to compute the radiative line driving. The
adopted surface abundance patterns are motivated using evolutionary tracks for rotating 
metal-poor stars. In sections~\ref{sec:windmodels} and~\ref{section:winddynamics} we 
present and discuss our wind models. Consulted evolutionary tracks are discussed in 
Sect.~\ref{sec:evoltracks}. The adopted grid is presented in Sect.~\ref{sec:grid}.

\subsection{Wind models}
\label{sec:windmodels}
To determine mass-loss rates, we use an iterative method,
as developed by \cite{2008A&A...492..493M} and \cite{2012muijresB}, in which {\sc isa-wind}
model atmospheres \citep{1993A&A...277..561D} are used to describe the model atmosphere and
{\sc mc-wind} models \citep{1997ApJ...477..792D,1999A&A...350..181V} 
to treat the wind dynamics. The latter uses a
Monte Carlo approach based on the method developed by \cite{1985ApJ...288..679A}. 
The advantageous aspects of this approach are that multiple photon scatterings are
taken into account, that excitation/ionisation changes of the gas are treated self-consistently,
and that one can quite easily dissect the relative contributions of individual
species to the line force. Essential
aspects of the wind dynamics are summarized in Sect.~\ref{section:winddynamics}. First, we
briefly summarize the main assumptions of the model atmospheres. 
For further details we refer the reader to the above references. 

We assume the wind to be homogeneous, spherically symmetric and stationary. Our model 
atmospheres extend from the base of the photosphere to approximately $20$ stellar radii.
For hydrogen, helium, carbon, nitrogen, oxygen, and silicon, we solve the ionization state 
and the occupation numbers of the levels in non-local thermodynamical equilibrium (NLTE). 
The other elements are treated using a modified nebular approximation. The radiative transfer 
in lines is calculated using the Sobolev method \citep{1960mes..book.....S}. 

To compute the line force we make
a selection of the strongest lines from the line-list in \cite{1995KurCD..23.....K}. 
This line list is essentially complete for the ions that play an important role in driving the 
winds studied here.
Transitions of the CNO elements of ionization species C\,{\sc v}, N\,{\sc vi} and O\,{\sc vi} 
or higher are not considered. These ions with one or few bound electrons have only
few lines in the part of the spectrum that coincides with the flux maximum of the
stellar radiation. As ionization of all these species depends on the radiation
field in the He\,{\sc ii} continuum above 54.4\,eV they would remain trace species 
in our models. Shocks in the outflows of massive stars (see \cite{2000ARA&A..38..613K}
for a discussion/review) may heat a fraction of the
gas to higher temperatures and produce non-thermal emission at soft X-ray wavelength. 
Such effects may increase the importance of transitions of these highly ionized species,
although it is unlikely that an associated increase in line force is significant. We
therefore do not expect that the lack of these lines has an important impact on our
mass loss predictions. Finally, we note that we miss 
part of the transitions of N\,{\sc vi}. However, for similar reasons 
we do not think that this affects our results in any notable way.
  
\subsection{Wind dynamics} 
\label{section:winddynamics}

The main challenge faced in determining the structure of a line driven stellar wind is the strong, but delicate,
interaction between radiation, line force, and atmospheric structure:
The radiation field and the density (or velocity) structure set the excitation/ionisation state of the gas. The line
force results from the structure and the state of the gas, while, in turn, the line force sets the structure, therefore
indirectly also the radiation field.
We try to tackle this problem through an iterative process. By assuming a velocity law and a mass-loss rate we can
determine the excitation/ionization structure. Next, the line force is calculated. Based on a fit to the 
line force the velocity law is re-determined. A model is considered converged 
once the fit parameters of the line force fit function and the mass-loss rate do not change significantly from one iteration step to the next. 

The procedure followed is based on
the method described by \cite{2008A&A...492..493M} and \cite{2012muijresB}, that improve on the treatment
developed by \cite{1997ApJ...477..792D} and \cite{1999A&A...350..181V}, and applied by 
\citet[e.g.][]{2000A&A...362..295V,2001A&A...369..574V}
Actually, the \citeauthor{2008A&A...492..493M} and \citeauthor{2012muijresB} studies present two approaches
to deal with the equation of motion.
In one solution, referred to as best-$\beta$ method, the velocity law is assumed to be a $\beta$-law
\citep{1978A&A....66..417L}. A fit function to the line force is constructed that is used to find the most
appropriate terminal velocity \vinf\ and measure for the rate of acceleration in the lower part of the wind,
expressed by $\beta$. In the other solution the fit function to the line force is used to numerically solve
the equation of motion, \textcolor{black}{referred to as the hydrodynamical method}. 
Both methods are compared in \cite{2012muijresB},  where it is found that the 
derived mass-loss rates do not differ more than about 25 percent and that the terminal wind
velocities agree within 20 percent.

The \textcolor{black}{hydrodynamical} method is dynamically consistent but is sensitive to the quality of the fit to the 
line force at the base of the wind, and therefore less stable. In the range of parameter space 
investigated in this article, we anticipate this to be more problematic than for the main sequence
O-stars studied by \citet{2012muijresB}.
For this reason we apply the best $\beta-$method from \cite{2008A&A...492..493M}.\\ 

The best $\beta$-method is, strictly speaking, not dynamically consistent.
This warrants caution. \citet{2012muijresB} discuss this issue in detail. They
identify the solution to be physically correct only if the
effective gravity $g_{\rm eff}$ -- i.e. the Newtonian gravity corrected for forces due to continuum
radiation pressure -- approximately balances the line force $g_{\rm line}$ at the location of
the sonic point $r_{\rm s}$, i.e.
\begin{equation}
   g_{\rm line}(r_{\rm s})  \simeq g_{\rm eff}(r_{\rm s})
   \label{eq:physical_sol_req}
\end{equation}
The sonic point corresponds to the radial distance where the flow velocity
reaches the local sound speed.
We compare the line force as simulated by the Monte Carlo code to the effective gravity 
and conclude that if the latter differs from the line force by more than 20 percent the wind can not
be driven by line radiation pressure alone. Phrased differently: in this case our model
does not drive a wind.

The condition Eq.\,(\ref{eq:physical_sol_req}) is more sophisticated than the one applied
by \cite{2009A&A...493..585K}. In their approach they apply a test to establish whether
a wind is driven that does not involve solving the equation of motion, and requires
specification of the wind mass-loss rate.

\subsection{Evolutionary tracks}
\label{sec:evoltracks}

Evolutionary tracks for rotating metal-poor massive stars have been computed by
\cite{2005A&A...443..643Y}, \cite{2006A&A...447..623M}, \cite{2006A&A...460..199Y} and 
\cite{2007A&A...461..571H}. The
set of models by \citeauthor{2005A&A...443..643Y} are for initially 12 to 60 \msun\ stars.
They span the metallicity range $Z = 10^{-5}$ to 
$2 \times 10^{-3}$ and cover initial equatorial rotation velocities between zero and 80 percent
of the Keplerian value. In these tracks,
transport of angular momentum is due to Eddington-Sweet circulations, shear instability,
Goldreich-Schubert-Fricke instability, and magnetic torques, and is approximated by diffusion.
In most cases, magnetic torques dominate, that are described in the context of the
Spruit-Tayler dynamo \citep{2002A&A...381..923S}, as explained by \citet{2005ApJ...626..350H}. The tracks by 
\cite{2007A&A...461..571H} are for initial masses between 20 and 85 \msun\ and cover metal 
fractions $Z = 10^{-8}$ to $0.02$ and
rotational velocities from zero to 800 \kmsec. Instabilities accounted for by 
\citeauthor{2007A&A...461..571H} are meridional circulation (treated as an advective process), 
and secular and dynamical shear instabilities (treated as diffusion).

\begin{figure}[t!]
\begin{center}
   \resizebox{9cm}{!}{\includegraphics[width=0.70\textwidth,angle=-90]{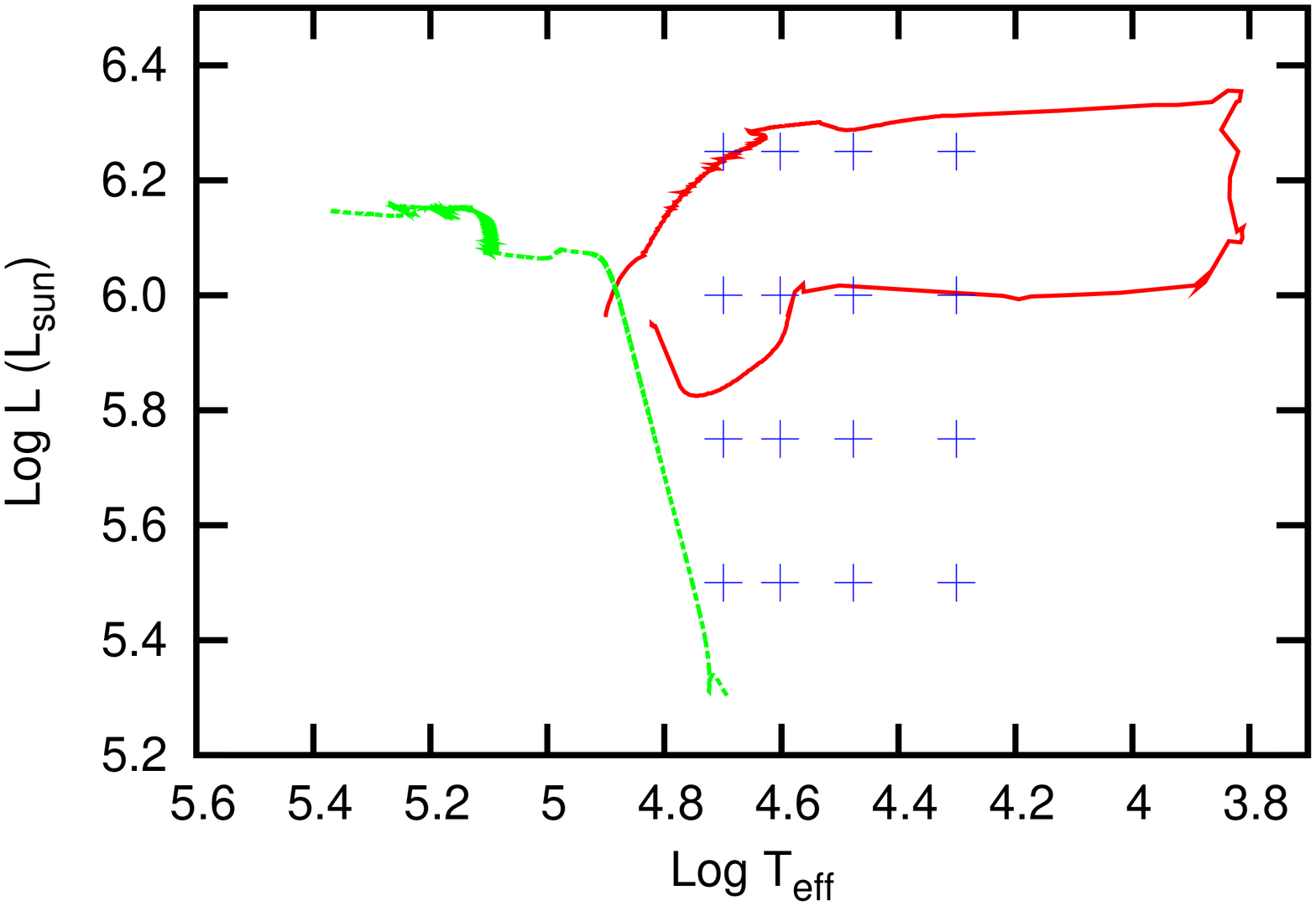}
              } 
   \caption{Evolutionary tracks for metal-poor stars. The green track is for
            $M_{\rm init} = 40\,\msun$, $Z = 10^{-5}$ and for a rotational velocity
            that is $v_{\rm rot} = 555$\,\kmsec\ \citep{2006A&A...460..199Y}. This 
            model evolves homogeneously. 
            The red track is for $M_{\rm init} = 85\,\msun$, $Z = 10^{-8}$ and
            $v_{\rm rot} = 800$\,\kmsec\ \citep{2007A&A...461..571H}. The blue pluses 
            denote ($L$,\teff) combinations
            for which mass-loss rates have been computed using abundance patterns
            that are typical for the surface composition of the evolutionary tracks.
            }
  \label{fig:HRD}
\end{center}
\end{figure}

Typical examples of the two sets of tracks are shown in Fig.~\ref{fig:HRD}. Interestingly,
the rapidly rotating tracks of \citet{2005A&A...443..643Y} and~\citet{2006A&A...460..199Y}
evolve more or less along the zero age main sequence (ZAMS) toward higher luminosities.
Hence, these stars remain hot and compact. The tracks of \citet{2007A&A...461..571H} evolve away from
the ZAMS, into the cool range of the Hertzsprung-Russell diagram. This striking difference
between the evolutionary tracks of the two groups is probably due to 
the inclusion of magnetic field in the Yoon et al. models. 
The different implementations of transport processes between the models might also have an 
impact and should be investigated properly but this is beyond the scope of this paper.

Different 
prescriptions are used to account for mass loss. \cite{2005A&A...443..643Y} and 
\cite{2006A&A...460..199Y} use results by 
\cite{1989A&A...219..205K}, assuming a metallicity dependence $\mdot \sim Z^{0.69}$ 
from \cite{2001A&A...369..574V}. For Wolf-Rayet phases they
use results by \cite{1995A&A...299..151H},
reduced by a factor 10, and a metallicity dependence as proposed by \cite{2005A&A...442..587V}. 
An {\em ad hoc} recipe for the effect of the increased CNO abundances on the
mass loss is used. 
\cite{2007A&A...461..571H} adopt the mass-loss recipes from \cite{2000A&A...360..227N} for 
Wolf Rayet phases. For O- and B-star phases, they use \cite{2001A&A...369..574V}. 
For stars cooler than 12\,500\,K  
they use \cite{1988A&AS...72..259D}. An artifact of the recipe of \citeauthor{1988A&AS...72..259D}
is that for luminous but cool stars (below $\sim$ 10\,kK - 20\,kK, depending on luminosity), outside of the
range of validity of the fitting function, a 20-term Chebychev polynomial, unrealistically 
large mass-loss rates are produced (of order $10^{-2}$ \msunyr). This rapidly strips
stars of their outer envelope, causing the tracks of \cite{2007A&A...461..571H} to become strongly enriched in this part of the HRD. A feedback occurs, as the metallicity 
associated with the enriched surface causes a high mass loss once they evolve back to higher
temperatures.
This implies that the surface enrichment in the blue-ward tracks of \citeauthor{2007A&A...461..571H} 
is strongly dependent on this (extreme) mass-loss history.

\subsection{Grid}
\label{sec:grid}

We limit the grid of models to the temperature range from 20\,kK to 50\,kK and
the luminosity interval from $10^{5.5}$ to $10^{6.25}$\,\lsun. The reason why we did not try
hotter models -- although predicted by \citet{2006A&A...460..199Y} -- is explained
below. Essentially, for such high temperatures in combination with a low iron group
metal content we do not find sufficient line driving to sustain a stellar wind. 
We encounter convergence problems of the model atmospheres for temperatures
close to 10\,000\,K because convection is not accounted for. To avoid convection
becoming an issue, we adopt 20\,kK as the low end of the investigated
temperature range.

For the surface abundance of carbon, nitrogen and oxygen combined, $Z_{\rm CNO}$, as well
as the relative abundances of these three elements we assumed values that
are typical -- but, deliberately, not tailored -- for the predictions of \citet{2006A&A...460..199Y} and 
\citet{2007A&A...461..571H}. Having said this, the abundance choices best fit
the 85\,\msun, $Z = 10^{-8}$ and $v_{\rm rot} = 800$\,\kmsec\ track for the case
of \citeauthor{2007A&A...461..571H} and the 40\,\msun, $Z = 10^{-5}$ and $v_{\rm rot}
= 555$\,\kmsec\ track for the \citeauthor{2006A&A...460..199Y} case. These are
the tracks that are shown in Fig.~\ref{fig:HRD}.
For the \citeauthor{2007A&A...461..571H} case we selected
two compositions, one that is characteristic for the start of the surfacing of CNO
processed material by rotational mixing (termed Case\,1) and one that is typical 
for the end of this mixing phase (termed Case\,2), but {\em before} the phase of extreme mass
loss as discussed in Sect.~\ref{sec:evoltracks}. The stars computed by
\citeauthor{2006A&A...460..199Y}  evolve on a chemically homogeneous track and are 
helium stars at the point they are enriched with hydrogen burning products. We adopt
a relative CNO abundance that is typical for the start of enrichment (termed Case\,4).
To separate the effects of the H/He ratio from CNO
abundance effects we introduce an intermediate Case\,3 in which the H/He ratio was
assumed to be equal to the H/He ratio from the \citeauthor{2007A&A...461..571H} track.
An overview of these abundance patterns is provided
in Table~\ref{table:grid2}.

For all these cases we adopt three values of $Z_{\rm CNO}$, i.e. 0.002, 0.02 and 0.04.
For abundances of other elements, we adopt the solar values
from \cite{1989GeCoA..53..197A} and scale them down to the metallicities as listed 
in Table\,\ref{table:grid2}. We note that these other elements do not contribute 
to the wind driving.

\begin{table}[t]
\begin{center}
\caption{Adopted abundance patterns: X and Y values are always consistent with
            the adopted $Z_{\rm CNO}$, but are given here for the case $Z_{\rm CNO} = 0.04$.   
            \label{table:grid2}}
\tiny
\begin{tabular}{llcccccc}
\hline\\[-9pt] \\[-5pt]
Case & Track type & X &  Y & C & N & O & $Z_{\rm iron}$ \\[1pt]
\hline\\[-7pt]  
\\[-6pt]
1 & Hirschi      & 0.56 & 0.40  & 42 \% & 38 \%  & 20 \% & $1 \times 10^{-8}$\\[1pt]
2 & Hirschi      & 0.56 & 0.40  & 55 \% & 25 \%  & 20 \% & $1 \times 10^{-8}$\\[1pt]
3 & Yoon et al. & 0.56 & 0.40  & 57 \% & 10 \%  & 33 \% & $1 \times 10^{-5}$\\[1pt]
4 & Yoon et al. & 0.05  & 0.91 & 57 \% & 10 \%  & 33 \% & $1 \times 10^{-5}$\\[2pt]
\hline
\end{tabular}
\end{center}
\normalsize
\end{table}

\section{Results}
\label{sec:Results}

\subsection{Mass-loss behavior}

The overall trends in the behavior of the predicted mass-loss rates are as follows:
For iron-group metallicities below \textcolor{black}{ $\sim$ 10$^{-4}$} (in line with \cite{2001A&A...369..574V})
and no primary enrichment of the atmosphere with carbon, nitrogen
and oxygen we find that we can not drive an outflow. Only for 
$Z_{\rm CNO} \sim 10^{-3}$ or higher, the CNO
elements can drive a stellar wind -- if circumstances are favorable. For temperatures 
of 50\,kK or higher the CNO elements are highly ionized. These ionic species -- such 
as C\,{\rm v}, N\,{\rm v} and O\,{\rm v} -- have relatively simple atomic level 
configurations, resulting in only few lines near the flux maximum of the emergent spectrum. 
The cumulative radiation pressure of these lines, supplemented with a contribution by 
hydrogen and helium lines that is diminishing for increasing temperature, is insufficient to
power a wind. Therefore, we do not predict stars hotter than $\sim$50\,kK to have a
line-driven outflow.

Table~\ref{table:grid} presents the \mdot, \vinf, and $\beta$ predictions for our grid.
For these models, characterized by a line driving that is on the brink of being sufficient 
to drive a wind, numerical uncertainties as well as the accuracy of the physical solution requirement 
Eq.~(\ref{eq:physical_sol_req}) may play a role in whether or not a solution is found. 
Typical uncertainties in the (iterative) numerical method itself are 0.1--0.2\,dex in the 
mass-loss rate and $\sim$30\,\% in the terminal velocity.
To avoid 
over-interpretation of individual models, we opt to present the results in terms of their overall
characteristics. Moreover, to provide a frame of reference we compare the computed mass-loss
rates with predictions based on the recipe of  \cite{2001A&A...369..574V}. These reference rates are
referred to as $\mdot_{\rm V}$ and are also provided in Table~\ref{table:grid}.

The \citeauthor{2001A&A...369..574V} recipe has been derived for a chemical composition of all 
elements apart from hydrogen and helium (i.e. the metallicity in the astrophysical context) that is 
scaled to the solar composition as given by \citet{1989GeCoA..53..197A}.  
We compare our mass loss \mdot\ with $\mdot_{\rm V}$ by identifying
$Z_{\rm CNO}$ of the models presented here as the metallicity $Z$ as intended by 
\citeauthor{2001A&A...369..574V}. Clearly, as a general trend we may expect that the 
\citeauthor{2001A&A...369..574V} recipe predicts larger mass-loss rates compared to the 
values computed here, as iron-group elements are more efficient in driving the wind through the 
sonic point than are lines of CNO only.

The $\mdot_{\rm V}$--recipe is valid for stars that are not too close to their Eddington
limit. In terms of the ratio of the radiation pressure on free electrons 
to the Newtonian gravity, i.e. $\Gamma_{\rm e} = g_{\rm e}/g_{\rm N}$, all $\mdot_{\rm V}$ 
predictions are
for $\Gamma_{\rm e} \la 0.4$. An extension of these predictions for stars close to their
Eddington limit by \cite{2011A&A...531A.132V} shows that for $\Gamma_{\rm e} \la 0.7$ the standard
$\mdot_{\rm V}$-recipe remains unaltered. The largest value of $\Gamma_{\rm e}$ in the
computations presented in this study are $\Gamma_{\rm e} \sim 0.5$ for models in which the 
luminosity $\log L/\lsun = 6.25$. 
We conclude that effects of approaching the Eddington limit do not complicate the
comparison of our results to the $\mdot_{\rm V}$ predictions.

\begin{figure}[t!]
     \resizebox{7.7cm}{!}{\includegraphics[width=0.70\textwidth,angle=0]{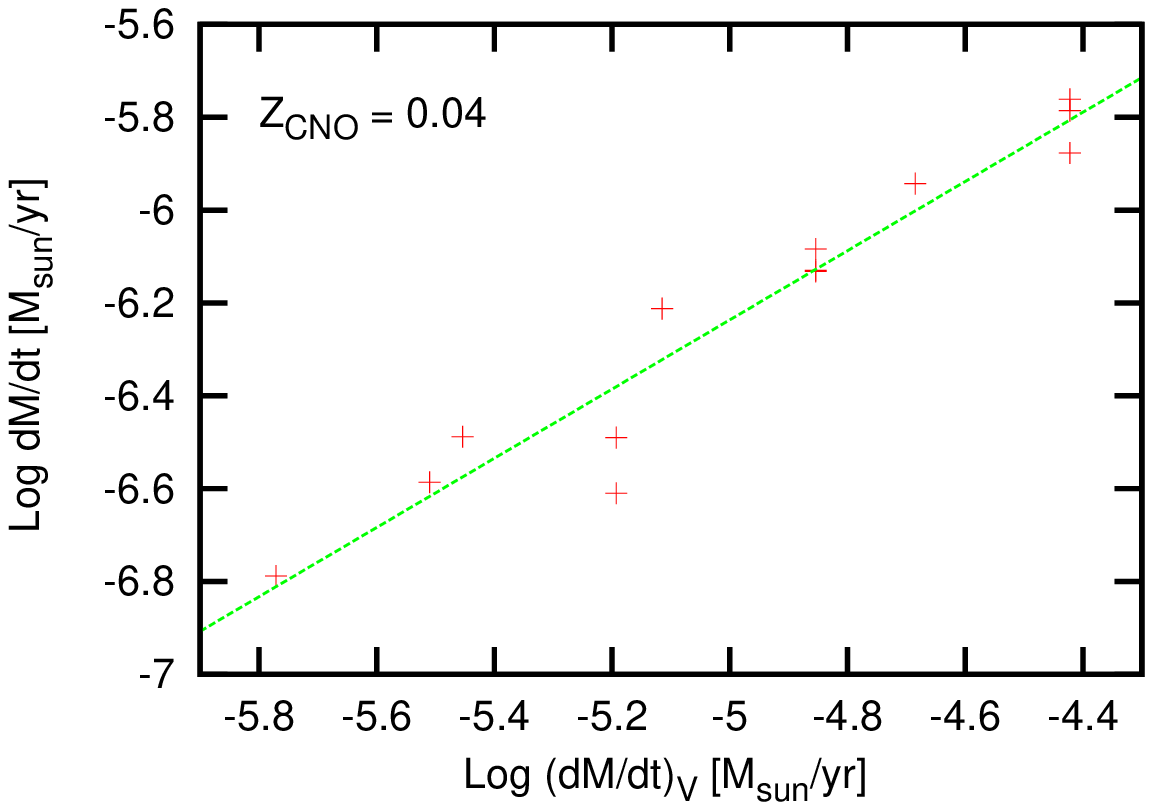}}
     \resizebox{7.7cm}{!}{\includegraphics[width=0.70\textwidth,angle=0]{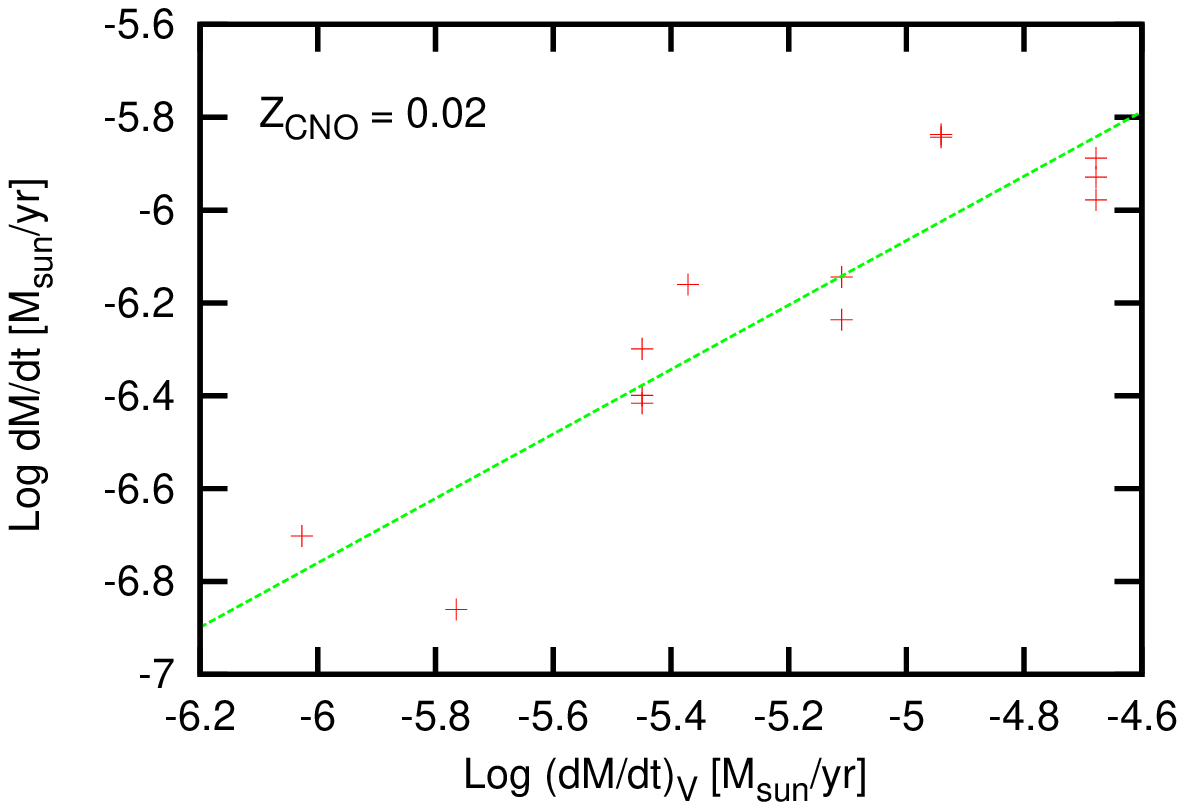}}     
     \resizebox{7.7cm}{!}{\includegraphics[width=0.70\textwidth,angle=0]{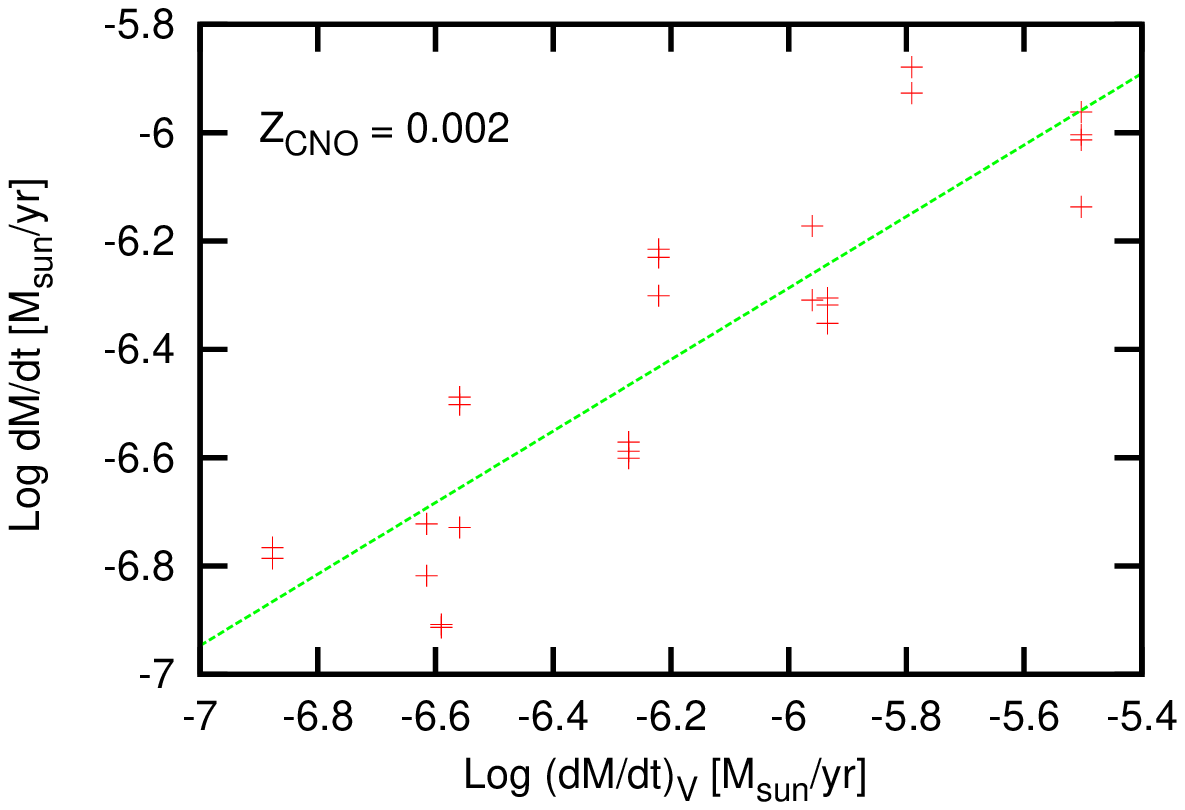}}     
  \caption{Derived mass loss rates versus predictions by \cite{2001A&A...369..574V} 
           (denoted $\mdot_{\rm V}$) for $Z=0.04$ (top panel), $Z=0.02$ (middle panel) and 
           $Z=0.002$ (bottom panel). The coefficients of the power law fits are given in 
           Table\,\ref{table:jorick}.}
           \label{fig:jorick}
\end{figure}

The outcome of the comparison is shown in Fig.~\ref{fig:jorick}, for the three different 
values of $Z_{\rm CNO}$, together with a power-law fit. The coefficients of these fits are 
listed in Table~\ref{table:jorick}. In all cases the behavior is approximately
\begin{equation}
   \mdot(Z_{\rm CNO}) = a \, \mdot_{\rm V}^{\alpha}(Z = Z_{\rm CNO} ) \sim \frac{1}{300} \mdot_{\rm V}^{0.7}(Z = Z_{\rm CNO}),
   \label{eq:mdot_mdot_vink}
\end{equation}
where the mass-loss rates are in units of \msunyr. Note that the metallicity $Z$ 
here concerns the CNO abundance, which is disjunct from the mass-loss $Z$ relations 
provided by \cite{2005A&A...442..587V} for WR stars where $Z$ referred to 
the Fe abundance.
We point out that the relation is only valid 
for the temperature range  20\,kK
to $\sim$50\,kK, and as long as the iron-group mass fraction is extremely 
low, i.e. below a few times 10$^{-4} Z_{\odot}$, in rough agreement with the 
mass-loss plateau of 10$^{-3} Z_{\odot}$ found by \citep{2005A&A...442..587V} 
for late-type WC stars. The error in the relation is typically a factor of three.

\begin{table}[b!]
\begin{center}
\caption{Fit parameters, expressing our predicted mass-loss rates for winds driven by
         carbon, oxygen and nitrogen lines, in terms of the mass-loss rates $\mdot_{\rm V}$
         as predicted by \cite{2001A&A...369..574V}: $\mdot = a \times \mdot_{\rm V}^{\alpha}$. 
         The coefficients are almost independent of $Z$, and can be summarized according
         to Eq.~(\ref{eq:mdot_mdot_vink}).
                         \label{table:jorick}
	     }
\small
\begin{tabular}{lcc}
\hline\\[-9pt] \hline\\[-7pt]
\multicolumn{3}{c}{Fit Parameters} \\[1pt]
$Z_{\rm CNO}$ & log $a$ & $\alpha$ \\[1pt]
\hline\\[-7pt]
0.04    & -2.51 $\pm$ 0.31  & 0.745 $\pm$  0.062   \\ [1.5pt]
0.02    & -2.60 $\pm$ 0.49  & 0.694 $\pm$  0.095   \\ [1.5pt]
0.002   & -2.33 $\pm$ 0.43  & 0.660 $\pm$  0.070   \\ [1.5pt]
\hline
\end{tabular}		         
\end{center}		         
\normalsize
\end{table}

Let us appraise some aspects of Eq.~(\ref{eq:mdot_mdot_vink}).  For a \citeauthor{2001A&A...369..574V}  
mass loss of $10^{-6}\,\msunyr$, our models imply a mass loss that is 0.3 dex lower for 
$Z_{\rm CNO} = 0.002$; 0.8 dex lower for $Z_{\rm CNO} = 0.02$, and 1.0 dex lower for $Z_{\rm CNO} 
= 0.04$. So, for progressively higher CNO enrichments the discrepancy with $\mdot_{\rm V}$ increases. 
This reflects the fact that the driving power of the CNO lines is topped once the relevant driving lines 
become saturated. 
An increase of the iron-group metal content to high metallicities of 0.02-0.04 does not suffer from 
this effect as most of the iron lines that drive the wind are relatively weak. Consequently,
the mass-loss rate of a wind in which the gas composition scales to the solar abundance pattern 
continues to rise with increasing metallicity. 
The fact that the main CNO driving lines are saturated in the $Z_{\rm CNO}$ range explored here 
implies that the absolute mass-loss rates of metal-poor stars enriched with primary CNO must be essentially independent of 
$Z_{\rm CNO}$. Indeed, this is what we find.

\subsection{Hydrogen rich versus helium rich stars} 

We find that overall it is more challenging for helium rich stars to drive a wind, if
all other stellar parameters (notably stellar mass) remain unaltered. We identify two
reasons for this. First, helium does not provide as many free electrons as 
hydrogen does. Therefore the contribution of the free electrons to the radiative force
is less. Second, the contribution of the helium lines to the radiative force is at maximum 
of the order of the contribution of the hydrogen lines but in many cases less. 
The latter effect is most relevant for the 20\,kK models, where the neutral hydrogen
fraction (although small in an absolute sense) is the highest.

The mass loss is found to be independent of H/He ratio. This could, however,
only be assessed for the case $Z_{\rm CNO} = 0.002$. We note that although
Cases 3 and 4 are representative for the homogeneously evolving tracks of 
\cite{2006A&A...460..199Y}, they need not be representative for strongly enriched
modestly rotating WNL (that may have $Z_{\rm N} > 0.15$) or WC stars (that may have 
$Z_{\rm C} > 0.50$). Models representative for such 'classical' type of Wolf-Rayet
stars have been computed by \cite{2005A&A...442..587V}.

\begin{figure}
   \begin{center}
       \resizebox{9cm}{!}{
     \includegraphics[width=0.90\textwidth,angle=0]{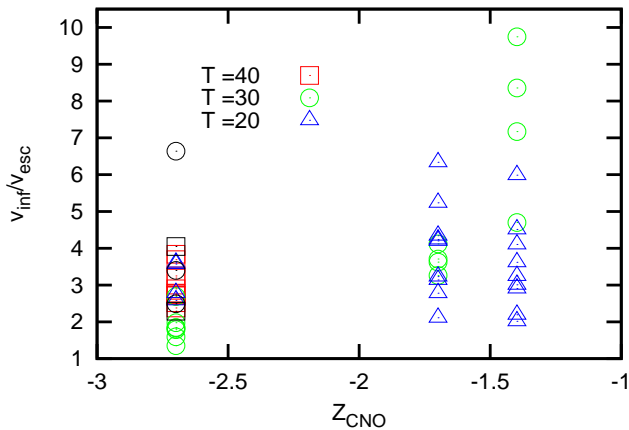}
      }
  \caption{Predicted terminal velocities \vinf\ for all models normalized to the effective 
           escape velocity at the stellar surface \vesc. Symbols and color codings
           indicate temperature; black symbols represent the H-poor models.
           Including the outliers towards high 
           \vinf/\vesc, the average of the H-rich models increases with $Z_{\rm CNO}$, from 2.6 to 3.9 to 4.7 for 
           $Z_{\rm CNO}$ = 0.002, 0.02 and 0.04, respectively.
	   }
  \label{fig:vinfvesc}
  \end{center}
\end{figure}

\subsection{Terminal velocity behavior}

Figure~\ref{fig:vinfvesc} shows the ratio of the terminal velocity to the effective surface escape
velocity as a function of $Z_{\rm CNO}$, for all converged models in our grid. As for the mass-loss
properties we also focus on the overall trends in the terminal flow characteristics. In line-driven
winds the nature of the driving lines (both in terms of chemical composition and ionization) is
reflected in the ratio $\vinf/\vesc$ \citep[see e.g.][]{1975ApJ...195..157C,2000ARA&A..38..613K}, 
therefore we discuss our results in these terms. 

For the H-rich stars in our grid (Cases 1 through 3) the trend is that $\vinf/\vesc$ is increasing 
with $Z_{\rm CNO}$. To be quantitative, the ratio increases from $2.7 \pm 0.7$ 
to $3.9 \pm 1.1$ to $4.7 \pm 2.4$ for $Z_{\rm CNO}$ = 0.002, 0.02 and 0.04, respectively. 
We note that the scatter in $\vinf/\vesc$ is rather large for the highest metallicity, and that for this case the models of 
20\,000\,K only yield a mean $\vinf/\vesc$ that is substantially lower, i.e. $3.5 \pm 1.2$.

The acceleration of the flow in the supersonic part of the wind is -- as in galactic O- and 
early B-stars -- controlled by strong lines of CNO \citep{1999A&A...350..181V,2000A&AS..141...23P}. In galactic OB-stars the observed ratio of 
terminal velocity to escape velocity is $\sim$2.6 for stars hotter than 22\,000\,K and about 
1.3 for cooler objects \citep{1995ApJ...455..269L}. 
Predictions of this ratio for normal O-stars by \cite{2012muijresB} for the best-$\beta$ method
(see Sect.~\ref{section:winddynamics}) yield \vinf\ values that are typically 
30\% higher than observed. Correcting for this, we still find higher outflows speeds.
The reasons why the CNO enriched metal-poor stars produce higher
terminal velocities than Galactic stars involves a combination of 
a lower mass-loss rate and a more efficient
driving in the supersonic part of the outflow, due to the high abundances of CNO. 
Relative changes in
the abundances of carbon, nitrogen and oxygen (as implied by Cases 1, 2 and 3) have a rather modest
effect on $\vinf$, given the uncertainty in the predictions of the terminal velocity ($\sim$30\%), this is not significant.

The converged He-rich models (Case 4) at 30\,000\,K tend to have higher $\vinf/\vesc$ ratios than do 
comparable H-rich models. In the He-rich case the continuum flux in the region where the main driving 
lines are located is larger, therefore these lines are more efficient, leading to larger \vinf.

\section{Discussion}
\label{sec:Discussion}

\subsection{Comparing to other studies}

Mass loss from luminous metal-free O-type stars enriched with primary CNO has been
studied by \cite{2009A&A...493..585K}, using stellar parameters from the evolutionary tracks for 
first stars, formed from metal-free gas clouds in the early Universe, by 
\citet{2001A&A...371..152M,2003A&A...399..617M}. 
A quantitative comparison of our results for metal-poor stars 
with those of \cite{2009A&A...493..585K} cannot be made because of 
the large differences in stellar parameters and surface $Z_{\rm CNO}$ 
abundances. Yet, we provide a rough comparison for a limited number of models. 
Accounting for the uncertainties introduced in their model grid spacing, 
\citeauthor{2009A&A...493..585K} found a similar \teff\ boundary for which stars
in similar luminosity ranges can drive winds: stars must be cooler 
than $\sim$60--50\,kK. Accepting a $\sim$10\% difference in the stellar mass,
we find similar mass-loss rates for our model with $Z_{\rm CNO} = 0.02$, 
\teff = 20\,kK and $\log L/\lsun = 5.9$ (see their model M500-3): the difference is 
$\sim$0.2\,dex. However, we obtain a terminal velocity that is 20--40 percent
lower, depending on the exact choice of chemical patterns.

We note that there are both similarities
and differences in our treatment of the stellar atmosphere and wind and those by
\citeauthor{2009A&A...493..585K}. In both methods the state of the gas is treated in
NLTE, whilst the Sobolev approximation is used to describe the transfer of radiation in
spectral lines. \citeauthor{2009A&A...493..585K}, however, treat the photosphere
-- from where the continuum radiation originates -- and wind separately. They adopt
an emergent flux taken from H-He model atmospheres, whereas we treat the photosphere
and wind in a self-consistent, unified way. Moreover, in our Monte Carlo treatment we
self-consistently account for multiple scattering effects, which are ignored in the
\citeauthor{2009A&A...493..585K} approach. For the type of models discussed here, these
differences are not likely to cause major differences. 

More important, however, is the difference pointed out in Sect.~\ref{section:winddynamics}.
\citeauthor{2009A&A...493..585K} apply a test in order to assess whether a wind can be 
sustained that is much simpler than the method applied here. The advantage of their 
criterion is that it allows one to {\em a priori} determine whether a star can drive 
a wind. However, such a test does not fully capture the feedback between line driving and
wind density, nor does it fulfill the requirement of dynamical consistency.
For the most luminous objects, we can compare models at both 30\,kK and 20\,kK,
again for $Z_{\rm CNO}$ = 0.02 (their models M999-1 and M999-2). Here we find
a mass-loss rate that is $\sim$0.3-0.7 dex lower, while terminal velocities
are lower by $\sim$20 percent. Here it should be noted that 
\citeauthor{2009A&A...493..585K} adopt a mass of 100\,\msun, while the tracks
on which we rely suggest $\sim$80\,\msun. 

This necessarily very limited comparison suggests that we predict roughly 
similar mass-loss rates for stars of initially 50--75\,\msun, but that for more 
massive stars, in the range of
75--100\,\msun\ our mass-loss rates are lower by a factor of a few.

\subsection{How to apply $\mdot(Z_{\rm CNO})$}

\citet{2001A&A...369..574V} point out that their recipe is valid for a scaled solar
metallicities as low as 1/30\,$Z_{\odot}$ for luminosities less than a million solar luminosities,
and 1/100\,$Z_{\odot}$ for more luminous objects. Their adopted solar values are from
\citet{1989GeCoA..53..197A}, in which $Z_{\rm \odot} = 0.019$. In exploring the line-driving
at their lowest metal contents, \citet{2001A&A...369..574V} already detect an increasing
relative importance of CNO. Moreover, they point out that the occurrence of the bi-stability
jump at about 35\,000\,K at very low metal content due to the recombination of ions of
a CNO element, i.e. C\,{\sc iv} to C\,{\sc iii} \citep{2000PhDT........45V}. 
We tested whether neglecting the up-turn
in mass-loss for stars cooler than 35\,000\,K in the present-day Universe in the 
$\mdot_{\rm V}$-recipe improved or degraded the correlation presented in Fig.~\ref{fig:jorick}
and Table~\ref{table:jorick}. This significantly degraded the correlation.

To provide a simple recipe to adjust
the \citeauthor{2001A&A...369..574V}-recipe for models that show primary enrichment, we
propose the heuristic formula:
\begin{equation}
   \mdot = \mdot(Z_{\rm CNO}) + \mdot_{\rm V}(Z_{\rm initial}),
   \label{eq:recipe}
\end{equation}
where $Z_{\rm initial}$ is the initial metal content of the star and $Z_{\rm CNO}$ is the primary CNO abundance. 
$\mdot(Z_{\rm CNO})$ is given by Eq.~(\ref{eq:mdot_mdot_vink}). In the $\mdot_{\rm V}$-recipe, the
bi-stability jump should still be applied.
Equation~\ref{eq:recipe} is valid for 
$\teff < 50\,{\rm kK}$. For hotter temperatures and abundances below $Z \sim$ a few $10^{-4}$ \citep{2001A&A...369..574V}, 
\textcolor{black}{we do not predict stars to lose mass} in a line-driven wind. The validity of the $\mdot_{\rm V}$ formula
extends to temperatures as low as 12\,500\,kK. The new $\mdot_{\rm CNO}$ predictions
have been computed down to 20\,000\,kK. A modest extrapolation may be applied, although
most certainly not below 15\,000\,kK. The highest $Z_{\rm CNO}$ abundance in our
computations is 0.04. Again, some extrapolation may be acceptable. However, for 
chemical compositions that are more typical for evolved Wolf-Rayet objects ($Z_{\rm CNO} \gtrsim 0.1$), we refer to
\citet{2005A&A...442..587V}. The lowest $Z_{\rm CNO}$ abundance in our computations
is 0.002. 
The minimum $Z_{\rm CNO}$ for which a wind may be driven is a function of luminosity,
proximity to the Eddington limit, and, to a lesser extent, detailed chemical abundance
pattern. However, the mass-loss rates associated with such low CNO abundances will
not impact the overall mass-loss and/or angular momentum loss of these stars
(see Sect.~\ref{sec:implications}). This
largely removes a practical need to define the limit below which metal-poor stars
may or may not drive a wind.

We point out that in terms of proximity to the Eddington limit, the stars investigated
show Eddington factors $\Gamma_{\rm e}$ up to $\sim$0.5. We therefore do not exclude
that metal-poor stars very close to their Eddington limit may still drive a wind
through line radiation pressure, as discussed by \citet{2002ApJ...577..389K,2002A&A...393..543V,
2005A&A...442..587V,2006A&A...446.1039K,2008A&A...482..945G,2011A&A...531A.132V}.

\subsection{Implications for the evolution of metal-poor stars}  
\label{sec:implications}  

Although the surfacing of CNO-cycle products may make the difference in whether
or not a metal-poor star can produce a line-driven wind or not, the actual
mass-loss rate such stars feature is only modest.
Absolute values of \mdot\ for the grid of stars presented here are in the range
$10^{-6}$--$10^{-7}$ \msunyr. Whilst substantial surface enrichment may only
occur after a significant part of the life of metal-poor massive stars, even when
simply assuming this would happen shortly after reaching the main-sequence the
cumulative mass-loss is only $\sim$1\,\msun\ for a 30\,\msun\ star and
$\sim$10\,\msun\ for a 80\,\msun\ star. So, at most a few percent of the initial
mass, but likely less than that. This is similar to the conclusions reached 
by \cite{2005A&A...442..587V} and \cite{2009A&A...493..585K}.

Purely by line-driven winds, metal-poor and initially metal-free stars may
inject at total of the order of $10^{-3}-10^{-1}$ \msun\ of CNO processed material 
into their (primordial) surroundings see \cite{2007A&A...461..571H} (table 3). 
This is substantially less than the enrichment with
CNO and other metals that is to be expected in the (pair-instability) supernova
that ends their lives. Still, it may be too early to conclude that a shift from
a top-heavy initial mass function (IMF), predicted in simulations of the first 
star-forming regions \citep{1999ApJ...527L...5B,2002ApJ...569..549N}, to a 
\citet{1955ApJ...121..161S} IMF, is due to dust grains formed from \emph{supernovae}
material. This outcome is based on the fact that a line-driven wind is the only mass-loss mechanism
a metal-poor or first star may suffer. Rotation close
to critical, perhaps in combination with approaching the Eddington limit 
\citep{2006A&A...447..623M} and/or pulsations \citep{2001ApJ...550..890B} may
boost the mass-loss rate of these objects. 
Mass-loss eruptions in a post-main
sequence phase, as for instance suffered by the massive star $\eta$\,Carinae 
\citep{1994PASP..106.1025H}, might play a role in the early Universe as well.
So, far the exact nature of these eruptions eludes us
\citep[see][]{2004ApJ...616..525O}, and possibly this mechanism does not depend on the chemical 
composition. If so, this type of mass-loss may have played a relatively
important role in the Universe early-on.

\section{Conclusions}
\label{sec:Conclusions}

We presented mass-loss calculations of metal-poor massive stars that, as a result of rotational
mixing, have become enriched at their surface with products of primary CNO burning. From 
predictions of the evolution of such stars we adopted several characteristic CNO enriched 
abundance patterns, and for these chemical patterns we performed tailored computations 
of the mass-loss rates driven by radiation pressure on spectral lines. 
The formal requirement to drive a wind in this study is
that the gravity balances the radiative forces at the sonic point of the flow.

The main conclusions we obtain are:

\begin{enumerate}

\item The mass-loss predictions $\mdot_{\rm V}$ \citep{2001A&A...369..574V} have not 
      been determined for massive metal-poor ($Z \la 1/100\,Z_{\odot}$) stars. 
      Applying this formula, and equalizing the CNO enriched chemical pattern 
      to $Z$, would overestimate the actual mass-loss rate.
      
\item On the basis of our new calculations, we provide a heuristic formula to estimate 
the mass-loss rates of such metal poor
stars, for the parameter range $L \ga 10^{5.5}\,\lsun$, $\teff \ga 15\,000$\,K and
$\Gamma_{\rm e} \la 0.5$, and we propose to apply this formula (Eq.~\ref{eq:recipe}) -- 
expressed as a scaling to the oft-used \cite{2001A&A...369..574V}-recipe -- in 
evolutionary predictions of such objects.

\item Metal-poor and surface CNO enriched massive stars hotter than 50\,000\,K do
      not feature a line-driven wind; highly ionized carbon, nitrogen and
      oxygen ions provide too few efficient driving lines. This result is similar to 
      that obtained for metal-free stars by \citet{2009A&A...493..585K}, within the uncertainties in the 
      \teff\ resolution of the two grid computations. It implies that homogeneously 
      evolving stars (which remain very hot during their entire life) do not have 
      line-driven winds in the initial metallicity range investigated here ($Z \lesssim 10^{-4}$).

\item In the range of $Z_{\rm CNO}=$ 0.002-0.04, the mass-loss rate is not a strong 
      function of the ratio of C to N to O.
            
\item The terminal wind velocities of CNO enriched metal-poor stars are higher than
      those of massive stars in the present-day Universe. For $Z_{\rm CNO} \sim$
      0.02--0.04 they are 25-50\% higher. The reasons for this behavior are a combination
      of the lower mass-loss rate and a more efficient driving in the supersonic
      part of the outflow, due to the high abundance of CNO.
      
\item The winds of massive very metal-poor stars ($Z \lesssim 10^{-4}$), whether they are CNO enriched or not, are so
      weak that they do not significantly impact the total mass and/or angular momentum loss
      during their evolution. If other mass-loss mechanisms, such as $\eta$\,Carinae type
      of mass eruptions, do not
      occur for such objects, their supernova explosions are expected to be responsible
      for the bulk of the early nucleo-synthetic enrichment. 
\end{enumerate}

\begin{acknowledgement}
JSV acknowledges financial support from the UK Science and Technologies Facility 
Council (STFC). 
RH acknowledges support from the World Premier International Research Center
Initiative (WPI Initiative), MEXT, Japan.
\end{acknowledgement}

\begin{table*}[h!]
\begin{center}
\caption{Adopted model parameters and predicted wind properties. Both the models for which a
         wind solution is achieved and those for which this is not the case are listed. The
         chemical compositions (column 7) correspond to the coding as given in 
         Table.~\ref{table:grid2}.
         Typical errors on the theoretical mass-loss rates are 0.1--0.2 dex; the errors on the
         terminal velocities are $\sim$30\%. \textcolor{black}{ For models in which the abundance pattern is followed by a semi colon
         the errors are larger.} The reference mass-loss rates $\mdot_{\rm V}$ are as 
         given by \cite{2001A&A...369..574V}, if we employ $\vinf = 2.6\,\vesc$.
         Beware that in order to compute $\mdot_{\rm V}$ we have set the value of
         $Z_{\rm CNO}$ equal to the metallicity value $Z$, which is not intended by the $\mdot_{\rm V}$--recipe, in which $Z$ refers to the 
	 initial iron abundance.
         Small changes in the effective escape velocity from the stellar surface (column 10)
         are due to abundance effects on $\Gamma_{\rm e}$. 
         \label{table:grid}
         }
\tiny
\begin{tabular}{rrrrlllrrrr}
\hline\\[-7.5pt] \hline\\[-6pt]
\multicolumn{7}{l}{Model Parameters} & \multicolumn{4}{l}{Wind properties} \\[1pt]
$L$    & $M$    &  $\teff$ & $R$ &  $Z_{\rm CNO}$ & $\mdot_{\rm V}$  & Abund. & $\mdot$ & $\vinf$ & \vesc\ &$\beta$ \\[1pt]  
$\log(\lsun)$ &  $\msun$ & kK   & $\rsun$ & & $\log(\msun/{\rm yr})$ & pattern & $\log(\msun/{\rm yr})$ & \kmsec\ & \kmsec\  
& \\[1pt]
\hline\\[-7pt]
5.50 & 30.0 & 50.0 & 7.5    & 0.04  &-5.496 & --       & --     & --   & --  & --    \\[1.5pt]
     &      &      &        & 0.02  &-5.752 & --       & --     & --   & --  & --    \\[1.5pt] 
     &      &      &        & 0.002 &-6.602 & --       & --     & --   & --  & --    \\[1.5pt]
     &      & 40.0 & 11.7   & 0.04  &-5.484 & --       & --     & --   & --  & --    \\[1.5pt]
     &      &      &        & 0.02  &-5.740 & --       & --     & --   & --  & --    \\[1.5pt] 
     &      &      &        & 0.002 &-6.590 & 3        & -6.908 & 2330 & 852 & 0.72  \\[1.5pt]  
     &      &      &        & 0.002 &-6.590 & 4        & -6.913 & 2060 & 900 & 0.64  \\[1.5pt]
     &      & 30.0 & 20.9   & 0.04  &-5.771 & 2        & -6.788 & 5406 & 647 & 1.12  \\[1.5pt]
     &      &      &        & 0.02  &-6.027 & 1        & -6.702 & 2100 & 644 & 0.78  \\[1.5pt] 
     &      &      &        & 0.002 &-6.877 & 1:       & -6.766 & 1150 & 644 & 0.66  \\[1.5pt] 
     &      &      &        &       &       & 3        & -6.786 & 1193 & 646 & 0.64  \\[1.5pt]
     &      & 20.0 & 47.0   & 0.04  &-5.510 & 2:       & -6.586 & 956  & 436 & 0.67  \\[1.5pt]
     &      &      &        & 0.02  &-5.765 & 1:       & -6.860 & 2750 & 434 & 1.23  \\[1.5pt] 
     &      &      &        & 0.002 &-6.615 & 1:       & -6.818 & 1550 & 434 & 0.84  \\[1.5pt]
     &      &      &        &       &       & 2:       & -6.722 & 1126 & 436 & 0.72  \\[1.5pt]     
5.75 & 45.0 & 50.0 & 10.0   & 0.04  &-5.179 & --       & --     & --   & --  & --    \\[1.5pt]
     &      &      &        & 0.02  &-5.435 & --       & --     & --   & --  & --    \\[1.5pt] 
     &      &      &        & 0.002 &-6.285 & --       & --     & --   & --  & --    \\[1.5pt]
     &      & 40.0 & 15.6   & 0.04  &-5.167 & --       & --     & --   & --  & --    \\[1.5pt]
     &      &      &        & 0.02  &-5.422 & --       & --     & --   & --  & --    \\[1.5pt] 
     &      &      &        & 0.002 &-6.272 & 1        & -6.571 & 2080 & 875 & 0.67  \\[1.5pt]
     &      &      &        &       &       & 2        & -6.588 & 2204 & 875 & 0.71  \\[1.5pt]
     &      &      &        &       &       & 3        & -6.601 & 2377 & 875 & 0.70  \\[1.5pt]
     &      & 30.0 & 27.8   & 0.04  &-5.454 & 3        & -6.488 & 3130 & 666 & 0.84  \\[1.5pt]
     &      &      &        & 0.02  &-5.709 & --       & --     & --   & --  & --    \\[1.5pt] 
     &      &      &        & 0.002 &-6.559 & 1        & -6.488 & 1060 & 665 & 0.66  \\[1.5pt]
     &      &      &        &       &       & 2:       & -6.502 & 900  & 665 & 0.60  \\[1.5pt] 
     &      &      &        &       &       & 4        & -6.729 & 4800 & 723 & 0.95  \\[1.5pt]		
     &      & 20.0 & 62.4   & 0.04  &-5.193 & 1        & -6.490 & 2030 & 450 & 0.86  \\[1.5pt]
     &      &      &        &       &       & 3        & -6.610 & 2694 & 450 & 1.09  \\[1.5pt]     
     &      &      &        & 0.02  &-5.449 & 1        & -6.399 & 1250 & 450 & 0.69  \\[1.5pt]
     &      &      &        &       &       & 2:       & -6.299 & 950  & 450 & 0.80  \\[1.5pt]
     &      &      &        &       &       & 3        & -6.416 & 1412 & 450 & 0.81  \\[1.5pt]
     &      &      &        & 0.002 &-6.299 & --       & --     & --   & --  & --    \\[1.5pt]
6.00 & 65.0 & 50.0 & 13.3   & 0.04  &-4.840 & --       & --     & --   & --  & --    \\[1.5pt]
     &      &      &        & 0.02  &-5.096 & --       & --     & --   & --  & --    \\[1.5pt] 
     &      &      &        & 0.002 &-5.946 & --       & --     & --   & --  & --    \\[1.5pt]
     &      & 40.0 & 20.8   & 0.04  &-4.828 & --       & --     & --   & --  & --    \\[1.5pt]
     &      &      &        & 0.02  &-5.084 & --       & --     & --   & --  & --    \\[1.5pt] 
     &      &      &        & 0.002 &-5.934 & 1        & -6.305 & 2500 & 863 & 0.72  \\[1.5pt]
     &      &      &        &       &       & 2        & -6.318 & 2400 & 863 & 0.71  \\[1.5pt]
     &      &      &        &       &       & 3        & -6.352 & 2752 & 863 & 0.75  \\[1.5pt]
     &      & 30.0 & 37.0   & 0.04  &-5.115 & 1        & -6.212 & 4740 & 661 & 1.06  \\[1.5pt]
     &      &      &        & 0.02  &-5.371 & 1:       & -6.160 & 2450 & 661 & 0.78  \\[1.5pt] 
     &      &      &        & 0.002 &-6.221 & 2        & -6.215 & 1305 & 660 & 0.64  \\[1.5pt]
     &      &      &        &       &       & 3:       & -6.230 & 1205 & 659 & 0.63  \\[1.5pt]
     &      &      &        &       &       & 4        & -6.301 & 2500 & 736 & 0.76  \\[1.5pt]   
     &      & 20.0 & 83.3   & 0.04  &-4.854 & 1        & -6.132 & 1625 & 449 & 0.80  \\[1.5pt]
     &      &      &        &       &       & 2:       & -6.130 & 1350 & 449 & 0.71  \\[1.5pt]   
     &      &      &        &       &       & 3:       & -6.084 & 905  & 449 & 0.65  \\[1.5pt]    
     &      &      &        & 0.02  &-5.110 & 1:       & -6.236 & 1950 & 449 & 0.96  \\[1.5pt] 
     &      &      &        &       &       & 2:       & -6.144 & 1450 & 449 & 0.82  \\[1.5pt]
     &      &      &        & 0.002 &-5.960 & 1        & -6.172 & 1250 & 449 & 0.76  \\[1.5pt]
     &      &      &        &       &       & 2:       & -6.309 & 1620 & 449 & 0.87  \\[1.5pt]
\end{tabular}		         
\end{center}		         
\normalsize
\end{table*}

\setcounter{table}{2}
\begin{table*}[t]
\begin{center}
\caption{Continued \dots 
         \label{table:grid1}}
\tiny
\begin{tabular}{rrrrlllrrrr}
\hline\\[-7.5pt] \hline\\[-6pt]
\multicolumn{7}{l}{Model Parameters} & \multicolumn{4}{l}{Wind properties} \\[1pt]
$L$    & $M$    &  $\teff$ & $R$ &  $Z_{\rm CNO}$ & $\mdot_{\rm V}$  & Abund. & $\mdot$ & $\vinf$ & \vesc\ &$\beta$ \\[1pt]  
$\log(\lsun)$ &  $\msun$ & kK   & $\rsun$ & & $\log(\msun/{\rm yr})$ & pattern & $\log(\msun/{\rm yr})$ & \kmsec\ & \kmsec\  
& \\[1pt]
\hline\\[-7pt]         
6.25 & 80.0 & 50.0 & 17.8   & 0.04  &-4.410 & --       & --     & --   & --  & --    \\[1.5pt]
     &      &      &        & 0.02  &-4.666 & --       & --     & --   & --  & --    \\[1.5pt] 
     &      &      &        & 0.002 &-5.516 & --       & --     & --   & --  & --    \\[1.5pt]
     &      & 40.0 & 27.8   & 0.04  &-4.398 & --       & --     & --   & --  & --    \\[1.5pt]
     &      &      &        & 0.02  &-4.653 & --       & --     & --   & --  & --    \\[1.5pt] 
     &      &      &        & 0.002 &-5.503 & 1        & -5.962 & 2310 & 711 & 0.76  \\[1.5pt]
     &      &      &        &       &       & 2        & -6.004 & 2616 & 709 & 0.80  \\[1.5pt]
     &      &      &        &       &       & 3:       & -6.013 & 2721 & 709 & 0.81  \\[1.5pt]
     &      &      &        &       &       & 4        & -6.137 & 3400 & 840 & 0.75  \\[1.5pt]  
     &      & 30.0 & 49.3   & 0.04  &-4.685 & 2        & -5.943 & 5400 & 554 & 1.16  \\[1.5pt]
     &      &      &        & 0.02  &-4.941 & 2:       & -5.837 & 2274 & 552 & 0.84  \\[1.5pt]
     &      &      &        &       &       & 3        & -5.843 & 2000 & 552 & 0.77  \\[1.5pt]
     &      &      &        & 0.002 &-5.791 & 2        & -5.879 & 1483 & 552 & 0.74  \\[1.5pt]
     &      &      &        &       &       & 4        & -5.927 & 1667 & 669 & 0.65  \\[1.5pt]     
     &      & 20.0 & 111.0  & 0.04  &-4.422 & 1        & -5.786 & 1110 & 382 & 0.74  \\[1.5pt]
     &      &      &        &       &       & 2:       & -5.877 & 1567 & 382 & 0.81  \\[1.5pt]
     &      &      &        &       &       & 3:       & -5.761 & 1240 & 382 & 0.81  \\[1.5pt]
     &      &      &        & 0.02  &-4.678 & 1        & -5.978 & 2000 & 382 & 0.91  \\[1.5pt]
     &      &      &        &       &       & 2        & -5.888 & 1615 & 381 & 0.93  \\[1.5pt] 
     &      &      &        &       &       & 3:       & -5.929 & 1600 & 381 & 0.83  \\[1.5pt] 
     &      &      &        & 0.002 &-5.528 & --       & --     & --   & --  & --    \\[1.5pt]
\hline
\end{tabular}		         
\end{center}		         
\normalsize
\end{table*}

\bibliographystyle{aa}
\bibliography{references.bib}

\end{document}